\documentclass[11pt, oneside]{article}   	% use "amsart" instead of "article" for AMSLaTeX format
\usepackage[T1]{fontenc}
\usepackage{lmodern}
\usepackage{geometry}                		% See geometry.pdf to learn the layout options. There are lots.
\usepackage{graphicx}				% Use pdf, png, jpg, or eps§ with pdflatex; use eps in DVI mode
\usepackage{amsmath}		
\usepackage{amssymb,amsfonts,amsthm}
\usepackage{mathtools,cool}
\usepackage{braket}

\usepackage[cal=boondoxo]{mathalfa}

\title{\textbf{\Large A Kerr/CFT Correspondence in Twistor Space}}
\author{Christian Kunz \\ \small{\textit{E-mail:} \href{mailto:kunz.christian.321@gmail.com}{kunz.christian.321@gmail.com}}}
\usepackage{plain}
\usepackage{hyperref}

\newcommand{\ud}{\mathrm{d}}
\numberwithin{equation}{section}
\allowdisplaybreaks

\begin{document}
  \maketitle
  \tableofcontents
  
\begin{abstract}
  Ingoing and outgoing principal null geodesics in Kerr spacetimes are characterized as part of parametrized families of strings in complex Kerr geometry and are associated with holomorphic curves in twistor space with help of the Kerr theorem. They are defined on 2-dimensional twistor manifolds, one for outgoing and one for ingoing principal null congruences, and are solutions of free twistor string models on these 2-dimensional twistor manifolds. Such a twistor string model implies a conformal field theory and, assuming the applicability of the Cardy formula, agreement with the Bekenstein-Hawking area law can be achieved depending on the effective central charge and temperature. A couple of (ambi)twistor string candidate models are examined.
\end{abstract}
\pagebreak

\section{Introduction}
     The use of complex world-lines (which can be interpreted as string worldsheets) in complex Kerr spacetimes has a long history, see \cite{Newman:2004, Burinskii:2013} and references therein, and also the contribution of Penrose to the book \cite{Wiltshire:2009} based on his paper \cite{Penrose:1997}. In particular, Burinskii\cite{Burinskii:2013} devoted many articles to expose string-like structures in complex Kerr-Schild geometry.\\
     
     A completely different line of investigation was to find holographic duals of Kerr spacetimes, stimulated by the AdS/CFT correspondence and termed Kerr/CFT correspondence. It started with extreme Kerr black holes \cite{Guica:2008} and was generalized to non-extreme cases \cite{Castro:2010, Haco:2018}. These results generated considerable interest.\\
     
     In this paper it is argued that twistor strings similar to ones of Burinskii can be regarded as special holomorphic solutions of free twistor string models on the 2-dimensional twistor manifolds \cite{Penrose:1986, Araneda:2019} that correspond to the two principal null congruences of the Kerr spacetime. Such a twistor string model implies a 2-dimensional conformal field theory (CFT), which is the reason for calling it a Kerr/CFT correspondence, although it is not based on holographic duality, but still might spawn some interest. Dealing with a CFT one can ask whether the Cardy formula \cite{Cardy:1986}  applies and agrees with the Bekenstein-Hawking area law. Obviously this depends on the specifics of how the twistor string is gauged. We examine the 4-dimensional ambitwistor string of Geyer, Lipstein, and Mason \cite{Geyer:2014} and an anomaly-free 4-dimensional twistor string found by the author \cite{Kunz:2020, Kunz_1:2020}. Both models show agreement with Bekenstein-Hawking, but the ambitwistor string is related to non-minimal conformal gravity whereas the other twistor string has more potential to actually describe the Kerr spacetime.\\
     
     In section 2 we review the twistor string approach of Burinskii and show that strings similar to ones he considered are solutions of free twistor string models defined on a pair of holomorphic 2-dimensional twistor manifolds that belong to the principal null congruences through the Kerr theorem \cite{Penrose:1986, Araneda:2019}.
     
     In section 3 we check whether the 4-dimensional ambitwistor model \cite{Geyer:2014} agrees with Bekenstein-Hawking using the Cardy formula and come to an affirmative answer if the central charge is made zero with help of the additional current algebra. On the other hand we argue that this model is not a good candidate because it is inherently conformal and cannot describe the Kerr spacetime.
     
     In section 4 we examine the twistor string of \cite{Kunz:2020, Kunz_1:2020}. It is shown that the model agrees with Bekenstein-Hawking like the previous ambitwistor string. On the other hand, because of the gauging of the translation group it has more potential to actually describe the Kerr spacetime. Further the model leads to the correct tree-level gravitational scattering amplitudes exactly because of the presence of ghost fields that come from the gauging of the translations.
     
     The last section 5 contains summary and discussion.\\
          
\section{Twistor Strings in Kerr Spacetime}
  \label{Strings}
  Any analytic shear-free null congruence can be characterized as a complex 2-parameter holomorphic family of $\alpha$-surfaces (see paragraph after the twistor form of the Kerr theorem 7.4.14 in \cite{Penrose:1986}). When applying this to the two principal null congruences in complexified Kerr spacetime, it leads to two 2-dimensional twistor manifolds endowed with a holomorphic structure, and similarly in dual twistor space for $\beta$-surfaces \cite{Penrose:1986, Araneda:2019}\footnote{Our convention for twistors and dual twistors is interchanged from the one used in \cite{Penrose:1986} and \cite{Araneda:2019}, as typically done for more than a decade in perturbative gauge theory, as already mentioned in \cite{Mason:2009}.}.\\
  
  More specifically the Kerr metric in terms of a tetrad in null coordinates can be written as \cite{Kerr:2008}
  \begin{align}
  &ds^2 = (d\zeta + Y dv)(d\overline{\zeta} + \overline{Y} dv) - (dv - \frac{hk}{(1 + Y \overline{Y})^2})k,\label{kerr-schild}\\
  &k = du + \overline{Y} d\zeta + Y d\overline{\zeta} + Y \overline{Y} dv.\nonumber
  \end{align}
  where
  \begin{align*}
  &u = z + t, &v = z - t, &&\zeta = x + iy,
  \end{align*}
  $k$ is a principal null congruence with $Y$ being a solution of the equation
  \begin{align}
  &\overline{\zeta}Y^2  + 2(z - ia)Y - \zeta = 0,
  \label{congruence}
  \end{align}
  in the corresponding Kerr theorem, and the multiplicative coefficient $h$ is 
  \begin{align}
  & h = 2m \text{Re}(2 Y_{\zeta}).
  \label{factor}
  \end{align}
  The roots of the equation are \cite{Kerr:2008}
  \begin{align*}
  &Y_1 = \frac{r\zeta}{(z+r)(r-ia)}, &2Y_{1,\zeta} = + \frac{r^3 + iarz}{r^4 + a^2 z^2},\\
  &Y_2 = \frac{r\zeta}{(z-r)(r+ia)}, &2Y_{2,\zeta} = - \frac{r^3 + iarz}{r^4 + a^2 z^2},
  \end{align*}
  where $r$ is a root of
  \begin{align}
  &\frac{x^2 + y^2}{r^2 + a^2} + \frac{z^2}{r^2} = 1.
  \label{radius}
  \end{align}
  Equation \eqref{congruence} can also be written in terms of a twistor $Z_a = (\mu^{\dot{\alpha}}, \lambda_{\alpha})$ as a quadratic equation in $Z_a$:
  \begin{align}
  &Z_a Q^{ab}Z_b = 0,  \text{\hspace{10pt} where \hspace{3pt}} &Y = \frac{\lambda_1}{\lambda_0}, && \mu^{\dot{\alpha}} = x^{\alpha \dot{\alpha}} \lambda_\alpha,  &&Q^{ab} = \begin{pmatrix}
  0 & ia & 0 & \frac{1}{2}\\
  ia & 0 & -\frac{1}{2} & 0\\
  0 & -\frac{1}{2} & 0 & 0\\
  \frac{1}{2} & 0 & 0 & 0
  \end{pmatrix}.
  \label{twistor-congruence}
  \end{align}
  This form of the equation was used by Penrose in \cite{Penrose:1986, Wiltshire:2009}. The solutions for $Z_a$, up to a multiplicative factor, determine the two 2-dimensional twistor manifolds. Equivalently, they are given by $Y_{1,2}$, together with the coordinates. The complex conjugate values $\overline{Y}_{1,2}$ determine the ones in the dual twistor space\footnote{Obviously, it is just a matter of convention which role twistors and dual twistor play here. They can be interchanged.}. $Y_{1,2}$ can be represented in a more suggestive manner in terms of Kerr coordinates $(u, r, \theta, \phi)$ defined by
  \begin{align*}
  &t = u \pm r , & x + iy = (r \mp ia) e^{i\phi} \sin \theta, && z = r \cos \theta,
  \end{align*}
  where $r$ still satisfies \eqref{radius}, but $u$ is different from the null coordinate $u$. In these coordinates $Y_{1,2}$ simply become
  \begin{align*}
  &Y_1= e^{i\phi} \tan \frac{\theta}{2}, &Y_2 = - e^{i\phi} \cot \frac{\theta}{2}.
  \end{align*}
  They describe, for constant $\phi$ and $\theta$, outgoing and ingoing principal null geodesics, with affine parameter $\pm r$ \footnote{In this representation, the geometric interpretation of the outgoing geodesics is more fittingly viewed as ingoing geodesics on the $r < 0$ sheet of extended Kerr spacetime\cite{Hawking:1973}.}.\\
  
  In complex spacetime these coordinates become complex and independent. The principal null geodesics can be represented as families of twistor curves or strings that are holomorphic in $w$ with $\mathfrak{Re} (w) \!=\! \pm r$, parametrized by $\phi$ and $\theta$ and projected onto the space $\mathbb{P}\mathbb{N}$  of null twistors ($\sim$ real spacetime). Alternatively, bundling up some of these parameters leads to two families of closed twistor strings holomorphic in $w$, one of circular form with $\mathfrak{Re} (w) \!=\! \phi \!\in\! [-\pi, \pi]$ and $\theta$ constant, and one sort of perpendicular to it with $\mathfrak{Re} (w) \!=\! \theta \!\in\! (-\pi, \pi)$ and $\phi$ constant. The first string family is periodic and the latter antiperiodic, reminiscent of Ramond (R) and Neveu-Schwarz (NS) type strings in superstring theory. The presence of an NS sector is a reflection of the ring singularity in Kerr spacetime and its expansion to extended Kerr spacetime with 2 sheets, one for $r > 0$ and one for $r \!<\! 0$ \cite{Hawking:1973}. Burinskii \cite{Burinskii:2004, Burinskii:2013} looked at similar (although not identical) types of strings, mainly in the context of the Kerr spinning particle. Of course, there are many more holomorphic twistor strings possible on these 2-dimensional twistor manifolds.\\
  
  All these holomorphic strings and the analogous ones in dual twistor space can be viewed as special solutions of a twistor string model on the cross product of each of the two 2-dimensional twistor manifolds in twistor space with the corresponding one in dual twistor space, with action
  \begin{align}
  S_0 = \frac{1}{2 \pi} \int\mathop{}\negthickspace{\ud^2\mathop{}\negthickspace z} \frac{1}{2} \left( W \cdot \overline\partial{Z} - Z \cdot \overline\partial{W} \right),
  \label{action_0}
  \end{align}
  where $Z$ denotes a twistor and $W$ a dual twistor, with components
 \begin{equation*}
 Z = \binom{\lambda_{\alpha}}{\mu^{\dot{\alpha}}}, \,\,\,\,W = \binom{\tilde{\mu}^{\alpha}}{\tilde{\lambda}_{\dot{\alpha}}}.
 \end{equation*}
  The presence of R and NS type strings means that the twistors should be worldsheet spinors. This model defines a Virasoro algebra with a central charge which depends on which symmetries are gauged. One can ask whether the Cardy formula \cite{Cardy:1986} for this model agrees with the Bekenstein-Hawking area law:
  \begin{equation*}
  S_{\text{Cardy}} = \frac{\pi^2}{3}|c_{\text{eff}}| T \stackrel{?}{=} S_{\text{BH}} = \frac{A}{4},
  \end{equation*}
  where $T$ is the temperature and $c_{\text{eff}} = c - 24\Delta_0$ is the effective central charge with $c$ the central charge and $\Delta_0$ the lowest eigenvalue of the $L_0$ Virasoro operator. Concerning the temperature, the question arises about its value. As the principal null geodesics generate the event horizon, we can take clues from the holographic Kerr/CFT correspondence \cite{Castro:2010, Haco:2018} and set
  \begin{equation}
  T = T_L + T_R = \frac{Mr_+}{2 \pi J},
  \label{temperature}
  \end{equation}
  where the ingoing and outgoing null congruences are considered to provide left and right temperature, respectively.
  If we could show $c = 12$ in units of $J$ and $\Delta_0 = 0$ like in the holographic Kerr/CFT correspondence, then $S_{\text{Cardy}} = 2 \pi M r_+ = S_{\text{BH}}$, and the Cardy formula indeed would agree with Bekenstein-Hawking. On the other hand, in the following sections we consider a couple of anomaly-free twistor string models with zero central charge where the agreement between Cardy and Bekenstein-Hawking is no longer obvious but still achievable as we will see.\\
  
  Another question is whether such a twistor string defined on the 2-dimensional twistor and dual twistor manifolds determines the Kerr spacetime, i.e. whether the mapping between spacetime and twistor string model is bidirectional. The action \eqref{action_0} by itself is clearly not sufficient because it can describe a spacetime only up to a conformal factor, i.e. such a spacetime will generally not satisfy the Einstein equations. The conformal scaling invariance needs to be broken in a specific manner to satisfy these equations.\\
 
\section{Ambitwistor String}
  \label{Ambitwistor}
 
 The action for the 4-dimensional ambitwistor string with no supersymmetry\footnote{Adding supersymmetry by changing twistors Z and W to supertwistors does not alter any result in this section.} is \cite{Geyer:2014}
  \begin{align}
  S_1 = \frac{1}{4 \pi} \int\mathop{}\negthickspace{\ud^2\mathop{}\negthickspace z}\left( W \cdot \overline\partial{Z} - Z \cdot \overline\partial{W} + a Z \cdot W \right) + S_j,
  \label{action_1}
  \end{align}
  where the scaling symmetry GL(1) of the twistor string model in section \ref{Strings} has been gauged, forcing the twistors and dual twistors to be ambitwistor pairs with zero GL(1) charge, and where an action for a worldsheet current algebra has been added, chosen such that the model is anomaly-free with zero central charge. In contrast to the Berkovits-Witten string\cite{Berkovits_1:2004} it is assumed that the twistor fields are worldsheet spinors as required from section \ref{Strings}. To get an idea about the spectrum one can look at the vertex operators \cite{Geyer:2014, Farrow:2018} which imply that the lowest $L_0$ eigenvalue is $\Delta_0 = 1$ in units of $J$ such that $|c_{\text{eff}}| = 24 \cdot J$. Therefore, we come to the conclusion that the Cardy formula gives twice the value of the Bekenstein-Hawking entropy. But one can argue that the model, originally defined on the full twistor space \cite{Geyer:2014}, only has one Virasoro algebra when viewed as restricted to the 2-dimensional twistor manifolds, not a \emph{left} and a \emph{right} one, such that the temperature should be averaged, and we end up with agreement between Cardy and Bekenstein-Hawking.\\
  
  Unfortunately, this ambitwistor model cannot be relevant for the Kerr spacetime because the MHV tree-level gravitational scattering amplitudes describe conformal gravity, similar to the Berkovits-Witten model \cite{Farrow:2018, Berkovits_1:2004}.\\

\section{Alternate Anomaly-free Twistor String}
  \label{TwistorString}
  The second model we consider has the action \cite{Kunz:2020, Kunz_1:2020, Kunz:2021}
   \begin{multline}
   S_2 = \frac{1}{2 \pi} \int\mathop{}\negthickspace{\ud^2\mathop{}\negthickspace z} \left\{
   \frac{1}{2} \sum_{i=1}^2 \left( W_i \cdot \overline\partial{Z_i} - Z_i \cdot \overline\partial{W_i} + \Theta_i \cdot \overline\partial{\Psi_i}  + \Psi_i \cdot \overline\partial{\Theta_i} \right)
   \right.\\
  + \sum_{i,j=1}^2 \lambda_i \cdot a_{1 i j} \cdot \tilde{\phi}_j
  + \sum_{i,j=1}^2 \tilde{\lambda}_i \cdot a_{2 i j} \cdot \phi_j
  + \sum_{i,j=1}^2\tilde{\lambda}_i \cdot b_{i j} \cdot \lambda_j
  + \left(W_1 W_2 \right) \mathop{}\negthickspace \vec{c} \, \cdotp \vec{\tau}  \mathop{}\negthickspace \begin{pmatrix} Z_1 \\ Z_2 \end{pmatrix}
  \left. \vphantom{\sum_i} \right\},
  \label{action}
  \end{multline}
  where for $i=1,2$ $Z_i$ are twistors, $W_i$ dual twistors, $\Psi_i$ fermionic bi-spinors, and $\Theta_i$ fermionic dual bi-spinors, with components
 \begin{equation*}
 Z_i = \binom{\lambda_{i \alpha}}{\mu_i^{\dot{\alpha}}}, W_i = \binom{\tilde{\mu}_i^{\alpha}}{\tilde{\lambda}_{i \dot{\alpha}}},
 \Psi_i = \binom{\phi_{i \alpha}}{\psi_i^{\dot{\alpha}}}, \Theta_i = \binom{\tilde{\psi}_i^{\alpha}}{\tilde{\phi}_{i \dot{\alpha}}},
 \end{equation*}
 and where again the strings are considered to be worldsheet spinors, partitioning them into an NS sector and a R sector. One unusual aspect of this model is the presence of two twistors instead of just one. Assuming that they represent two intersecting $\alpha$-curves, due to the incidence relations $\mu_i^{\dot{\alpha}} = x^{\alpha \dot{\alpha}} \lambda_{i \alpha}$, then they are like two rays intersecting the celestial sphere over $x$ and are determined only up to a complex SU(2)\footnote{SU(2) $\sim$ SL(2, $\mathbb{C}$) as complex Lie algebra} symmetry operation. This is the reason for the gauged SU(2) symmetry between the two twistors in \eqref{action}, with Lagrange multiplier field $\vec{c}$ \cite{Kunz:2020}\footnote{$\vec{\tau}$ denote the Pauli matrices.}. Further, Isenberg \& Yasskin \cite{Isenberg:1986} showed that the twistor space and its dual are in natural correspondence with teleparallel spacetimes, which typically are modeled by gauging the translation group, suggesting the gauging of the translations on the worldsheet as well, with Lagrange multiplier field $b_{i j}$ \cite{Kunz:2020}. Finally, fermionic spinors have been added to the model, with worldsheet supersymmetries based on gauging of supertranslations, with Lagrange multipliers $a_{1 i j}$ and $a_{2 i j}$ \cite{Kunz:2020}. After performing BRST quantization, this model is anomaly-free (the BRST charge is nilpotent) in self-contained fashion, without need of an additional worldsheet current algebra \cite{Kunz:2020}.\\
 
 Analysis of the spectrum \cite{Kunz_1:2020} shows that the gauging of the translation symmetry reduces the number of complex degrees of freedom for each twistor from three to two. Therefore, we can choose a gauge of the SU(2) symmetry that assigns each twistor to one of the two 2-dimensional twistor manifolds in consistent manner, and the dual twistors in analogous way. This choice will reveal itself as very convenient for going back to the original spacetime.\\
 
 Like in the case of the ambitwistor string $\Delta_0 = 1$, and to make the Cardy formula agree with Bekenstein-Hawking, it needs to be adjusted to half the value. This makes absolute sense, considering that the single Virasoro algebra has two contributions, one from each of the two twistors, but gauged in such a way that it is to be considered a single contribution.\\
 
 Can this model actually represent the Kerr spacetime? It has been shown that it provides the expected tree-level gravitational scattering amplitudes in the NS sector \cite{Kunz:2020, Kunz_1:2020}, and that this happens precisely because the contractions between ghost and antighost fields arising from the gauging of the translation group are ensuring that only connected trees (or equivalently trees without loops) are allowed amongst the contractions in the worldsheet correlator of gravitational scattering \cite{Adamo_1:2012}. Also, by being able to associate each twistor of the model with a particular one of the two 2-dimensional twistor manifolds we can keep track of these manifolds separately. And for Kerr spacetimes twistors on these manifolds fulfill an homogeneous quadratic equation of the form \eqref{twistor-congruence} which can be evaluated and solved for $Y_{1,2}$, and allows to calculate the conformal factor \eqref{factor}, up to the mass factor. So, indeed, we get back the original Kerr spacetime, up to the mass which, of course, is nowhere to be found in twistor space.\\
 
This might look like cheating, inserting the knowledge that the spacetime was Kerr to begin with. On the other hand, one should note that the whole construction of the two 2-dimensional twistor manifolds can be generalized to any Petrov type D spacetime with two gravitational shear-free principal null congruences, the main difference to Kerr being that $Z_a Q^{ab}Z_b$ in \eqref{twistor-congruence} is replaced by a general holomorphic homogeneous function in $Z_a$. Knowing the two twistor manifolds Araneda\cite{Araneda:2019} showed that with certain assumptions that for instance are satisfied by Kerr-(A)dS and Kerr-Newman-(A)dS spacetimes, the original Petrov type D spacetime can be recovered via a conformally K{\"a}hler structure\cite{Dunajski:2009}, but only up to a conformal factor which would need to be fixed to ensure the validity of Einstein equations with or without cosmological constant. Whether this twistor string model, with help of the gauged translation symmetry, determines the correct conformal factor in this more general case is an open issue. For more discussion on this topic, in particular on applying the Isenberg \& Yasskin programme \cite{Isenberg:1986} for general spacetimes to this model, see \cite{Kunz:2021}.\\
  
\section{Summary and Discussion}
  \label{Discussion}
  In this paper we presented a non-holographic correspondence between the Kerr spacetime and a CFT in twistor space together with its dual based on twistor string models that are defined on a couple of holomorphic 2-dimensional twistor manifolds, one for each principal null congruence, and have holomorphic strings of principal null geodesics as classical solutions. We examined a couple of gauged twistor string models to see whether the Cardy formula leads to the same entropy as the Bekenstein-Hawking area law and whether they are candidates for representing the Kerr spacetime. Both models could be made to agree between Cardy and Bekenstein-Hawking, but the first model, the 4-dimensional ambitwistor string \cite{Geyer:2014}, relates to non-minimal conformal gravity whereas the second twistor string model \cite{Kunz:2020, Kunz_1:2020} provides the correct Einstein gravitational scattering amplitudes by gauging the translation symmetry on the worldsheet and actually describes the Kerr spacetime with help of the congruence equations \eqref{twistor-congruence} and \eqref{factor}. In order for this model to go from a Kerr spacetime to more general Petrov type D spacetimes, the effect of the various worldsheet gauge symmetries, especially of the translation symmetry, needs to be investigated more thoroughly.\\
  
  If the second model is actually a viable theory, it makes some interesting predictions. There are quite a few exotic particles in the spectrum never seen before \cite{Kunz_1:2020}. The spin 2 and $\frac{3}{2}$ excitations in the R sector can easily become massive at low energies by picking up corresponding spin 0,1, and $\frac{1}{2}$ excitations, but in the NS sector there are no obvious lower spin excitations available to make the graviton, gravitinos, and vector particles massive. This would mean that the NS sector only contains gravitationally interacting massless particles which should be detectable in gravitational waves (theoretically but experimentally with extreme difficulty), and the spin $\frac{1}{2}$ matter content is exclusively delegated to the R sector, providing an explanation why semi-classical quantum field theory on curved spaces can be so successful \cite{Parker:2009} and why gravitation is at a so different and weaker strength than the other interactions. And if the low energy limit of the theory exists, it is a modified gravity model with both hot and cold dark matter. A lot of details would need to be worked out.\\

\bibliography{TwistorString}
\end{document}